\documentclass[sigconf,screen,authorversion,nonacm]{acmart}
\AtBeginDocument{%
  }

\usepackage{multirow}
\usepackage{booktabs}
\usepackage{tabularx}
\usepackage{caption}
\usepackage{subcaption}
\usepackage{balance}
\usepackage{graphicx}

\usepackage{soul}
\usepackage[dvipsnames]{xcolor}

\usepackage{soul}

\setcopyright{acmlicensed}
\copyrightyear{2018}
\acmYear{2018}
\acmDOI{XXXXXXX.XXXXXXX}
\acmConference[Conference acronym 'XX]{Make sure to enter the correct
  conference title from your rights confirmation email}{June 03--05,
  2018}{Woodstock, NY}
\acmISBN{978-1-4503-XXXX-X/2018/06}

\begin{document}

\title{Algorithmic Audit of Personalisation Drift in Polarising Topics on TikTok}

\author{Branislav Pecher}
\affiliation{%
  \institution{Kempelen Institute of Intelligent Technologies}
  \city{Bratislava}
  \country{Slovakia}}
\email{branislav.pecher@kinit.sk}
\orcid{0000-0003-0344-8620}

\author{Adrian Bindas}
\affiliation{%
  \institution{Kempelen Institute of Intelligent Technologies}
  \city{Bratislava}
  \country{Slovakia}}
\email{adrian.bindas@kinit.sk}
\orcid{0009-0002-8796-6475}

\author{Jan Jakubcik}
\affiliation{%
  \institution{Kempelen Institute of Intelligent Technologies}
  \city{Bratislava}
  \country{Slovakia}}
\email{jan.jakubcik@kinit.sk}
\orcid{0009-0009-5162-2858}

\author{Matus Tuna}
\affiliation{%
  \institution{Kempelen Institute of Intelligent Technologies}
  \city{Bratislava}
  \country{Slovakia}}
\email{matus.tuna@kinit.sk}
\orcid{0009-0001-2062-8255}

\author{Matus Tibensky}
\affiliation{%
  \institution{Kempelen Institute of Intelligent Technologies}
  \city{Bratislava}
  \country{Slovakia}}
\email{matus.tibensky@kinit.sk}
\orcid{0000-0001-9928-3474}

\author{Simon Liska}
\affiliation{%
  \institution{Kempelen Institute of Intelligent Technologies}
  \city{Bratislava}
  \country{Slovakia}}
\email{simon.liska@kinit.sk}
\orcid{0009-0008-0873-9797}

\author{Peter Sakalik}
\affiliation{%
  \institution{Kempelen Institute of Intelligent Technologies}
  \city{Bratislava}
  \country{Slovakia}}
\email{peter.sakalik@kinit.sk}
\orcid{0009-0005-8959-3838}

\author{Andrej Suty}
\affiliation{%
  \institution{Kempelen Institute of Intelligent Technologies}
  \city{Bratislava}
  \country{Slovakia}}
\email{andrej.suty@kinit.sk}
\orcid{0009-0004-0787-7718}

\author{Matej Mosnar}
\affiliation{%
  \institution{Kempelen Institute of Intelligent Technologies}
  \city{Bratislava}
  \country{Slovakia}}
\email{matej.mosnar@kinit.sk}
\orcid{0009-0009-2466-7225}

\author{Filip Hossner}
\affiliation{%
  \institution{Kempelen Institute of Intelligent Technologies}
  \city{Bratislava}
  \country{Slovakia}}
\email{filip.hossner@kinit.sk}
\orcid{0009-0005-6630-0158}

\author{Ivan Srba}
\affiliation{%
  \institution{Kempelen Institute of Intelligent Technologies}
  \city{Bratislava}
  \country{Slovakia}}
\email{ivan.srba@kinit.sk}
\orcid{0000-0003-3511-5337}

\renewcommand{\shortauthors}{Pecher et al.}

\begin{abstract}
Social media platforms have become an integral part of everyday life, serving as a primary source of news and information for many users. These platforms increasingly rely on personalised recommendation systems that shape what users see and engage with. While these systems are optimised for engagement, concerns have emerged that they may also drive users toward more polarised perspectives, particularly in contested domains such as politics, climate change, vaccines, and conspiracy theories. In this paper, we present an algorithmic audit of personalisation drift on TikTok in these polarising topics. Using controlled accounts designed to simulate users with interests aligned with or opposed to different polarising topics, we systematically measure the extent to which TikTok steers content exposure toward specific topics and polarities over time. Specifically, we investigated: 1) a preference-aligned drift (showing a strong personalisation towards user interests), 2) a polarisation-topic drift (showing a strong neutralising effect for misinformation-themed topics, and a high preference and reinforcement of interest of US politic topic); and 3) a polarisation-stance drift (showing a preference of oppose stance towards US politics topic and a general reinforcement of users' stance by recommending items aligned with their stance towards polarising topics). Overall, our findings provide evidence that recommendation trajectories differ markedly across topics, with some pathways amplifying polarised viewpoints more strongly than others and offer insights for platform governance, transparency and user awareness.
\end{abstract}

\ccsdesc[500]{Information systems~Personalization}
\ccsdesc[300]{Information systems~Content ranking}
\ccsdesc[300]{Human-centered computing~Human computer interaction (HCI)}

\keywords{algorithmic audit; social media platform; sockpuppeting; personalisation; TikTok; polarising topics}


\maketitle

\section{Introduction}

Social media platforms increasingly rely on personalised and algorithmically curated recommendation systems that shape what users see based on their past interactions~\cite{klug2021trickplease, zannettou2024analyzing}. While such systems can enhance user engagement, reduce information overload, and tailor content to personal tastes, they may also exhibit problematic algorithmic behaviour~\cite{bandy2021survey}, such as amplification of harmful content. 

To shed more light on such undesired behaviour, recent studies employed \textit{algorithmic audits}, a technique that can be defined as an ``assessments of the algorithm’s negative impact on the rights and interests of stakeholders, with a corresponding identification of situations and/or features of the algorithm that give rise to these negative impacts''~\cite{10.1177/2053951720983865}. Due to the black-box nature of social media recommender systems, algorithmic audits explore their properties behaviourally: user interactions with an algorithm are simulated (e.g., content visits), and observed responses (e.g., recommended items) are examined for the presence of the audited phenomenon. Previous algorithmic studies have found, for example, that social media recommender systems tend to enclose users in filter bubbles, which may lead to the proliferation of problematic or misleading content, such as misinformation~\cite{tomlein_audit_2021, urman2024mapping, bandy2021survey}.

Besides spreading harmful content, there is a mounting concern with \textit{personalisation drift} on these platforms, where the recommender systems gradually steer users towards more polarised or extreme contents and stances~\cite{Spinelli2020, Ribeiro2020}. Such a tendency is particularly problematic for highly \textit{polarising topics}, such as politics, conspiracy theories or misinformation, where the high exposure can distort public understanding and deepen social divisions.

The existing works studying personalisation drift, focused primarily on political topics, especially in the context of elections across different countries. They revealed a wide range of problematic algorithmic behaviours, like amplification of right-leaning content for new, existing users, and even for users with left-leaning tendencies~\cite{ye2025auditing, ibrahim2025tiktok}, radicalisation and polarisation~\cite{shin2024algorithms}, or rewarding divisiveness, negative emotions and negative partisanship~\cite{solovev2025tiktok, ibrahim2025tiktok}. However, other studies have found opposing findings, where the recommendation either pushes users towards the same stance~\cite{ye2025auditing} or tends to rapidly shift users towards more mainstream content and non-polarising topics that are often more popular~\cite{ledwich2022radical, cakmak2025investigating}. It can be concluded that a systematic study of this phenomenon, which would provide focused insight, is currently missing.

In this paper, we present \textbf{an algorithmic audit of personalisation drift in polarising topics on TikTok}. We systematically measure how the platforms' recommender systems alter the content users are exposed to over time: 1) for different polarising as well as a neutral topic; and 2) with a supporting vs. opposing stance towards such topics. Namely, we compare this drift across 5 topics: \textit{flatearth conspiracy}, \textit{US politics}, \textit{vaccination}, \textit{climate change} and a neutral topic of \textit{cooking}. These topics were chosen to reflect different angles that are dividing society and are present on TikTok.

We look at three different types of personalisation drift, which we understand as a gradual change over time in the proportion of videos with specific characteristics displayed in the user's feed:
\begin{itemize}
    \item \textbf{Preference-aligned drift} that specifies the gradual change in ratio between videos of interest (i.e., a combination of neutral and polarising videos) and non-related videos.
    \item \textbf{Polarisation-topic drift} that specifies the gradual change in ratio between the polarising and neutral videos that are of interest for the user.
    \item \textbf{Polarisation-stance drift} that specifies the gradual change in ratio between different polarities of videos in the polarising topics.
\end{itemize}

For the purpose of this audit, we build controlled (sockpuppet) accounts that simulate user characteristics, behaviour, and interests for the different stances on these polarising topics\footnote{To support replicability, the anonymised collected data with their predicted annotations are available for research purposes at Zenodo upon request: \url{https://doi.org/10.5281/zenodo.19144520}; the code used for data analysis is publicly available as a Github repository: \url{https://github.com/kinit-sk/ai-auditology-personalisation-drift-tiktok}.}.
We chose TikTok as a platform due to its popularity (especially among the younger generations) and the already observed problems on this platform, even as part of the DSA audit reports\footnote{A European legislation, Digital Services Act (DSA), requires algorithms of online platforms to comply with specific obligations concerning algorithmic transparency, user protection and privacy. To verify compliance with these requirements, DSA mandates platforms to undergo independent audits.}~\cite{solarova2026beyondthecheckbox}.

Overall, we focus on answering the following research questions:
\begin{itemize}
    \item \textbf{(RQ1)} How does the preference-aligned drift and polarisation-topic drift exhibit across different topics of interest?
    \item \textbf{(RQ2)} How does polarisation-stance drift exhibit across topics of interest?
\end{itemize}

Within the first research question, we are primarily interested in the general personalisation behaviour of the TikTok recommender system (preference-aligned drift) and in whether the TikTok recommender system promotes polarising topics (characterised by divisiveness and negative emotions) in the same manner as popular neutral topics (polarisation-topic drift). Within the second research question, we are mostly interested in whether there is overall drift towards a specific stance in the topics (e.g., left or right-leaning videos in the case of US politics), or whether both stances are treated the same way, either by pushing users towards the extremes or by providing opposing stances.

The main contributions of this work are as follows:
\begin{enumerate}
    \item This paper makes a timely and socially relevant contribution by conducting a systematic algorithmic audit of TikTok’s recommendation system on sensitive and polarising topics, such as climate change and vaccines. In contrast to the existing works, it goes beyond politics and address also topics characterised by the presence of conspiracies and misinformation.
    \item The proposed audit methodology based on large language model–driven sockpuppet accounts allows for more realistic simulation of user behaviour and content annotation, and more extensive study using 68 user accounts. Comparison with a neutral topic further enables a clearer isolation of personalisation effects attributable to the platform’s recommender system. Various experimental setups finally allow us to study and distinguish three different types of personalisation drifts: a preference-aligned, a polarisation-topic, and a polarisation-stance drift. Such differentiation enables a clear separation of the overall personalisation effect, personalisation at the level of a polarising topic, and at the level of stance to such a topic.
    \item The obtained findings provide novel up-to-date insights into how TikTok recommender system address different polarising topics. First, we observe a positive behaviour -- personalisation for a neutral topic is much stronger than for polarising ones (except US politics). We also observe a strong neutralising effect for some polarising topics (climate change or vaccines). These findings confirm those from the previous works~\cite{ledwich2022radical, cakmak2025investigating}, which found that the TikTok rapidly shifts users to mainstream and popular, yet safe, topics. Second, for the topic of US politics, which receives strong attention in TikTok recommender, we observe a strong tendency and gradual drift towards videos from oppose stance (which may be the effect of events happening in the country). This is in contrast with the previous studies that observed that TikTok would push users towards right-leaning content~\cite{ye2025auditing, ibrahim2025tiktok, shin2024algorithms}.
\end{enumerate}

\section{Related Work}
Algorithmic audits have recently become a popular tool for investigating potentially harmful behaviours of the recommender systems on social media platforms~\cite{urman2024mapping, bandy2021survey, sandvig2014auditing}. Due to the increasing strength of the recommender systems utilised by social media, audits primarily focus on the distortion and disparities in content delivery for different users~\cite{urman2024mapping}. This includes a better understanding of the recommender system itself and the factors that affect it, such as gender, age, location or the implicit and explicit user interactions~\cite{mosnar2025revisiting, boeker2022empirical, kaplan2024comprehensively, vombatkere2024tiktok, robertson2018auditing, evans2023google, thorson2021algorithmic, Hussein2020}. Besides better understanding, multiple studies focus on the proliferation of the harmful content~\cite{Spinelli2020, Ribeiro2020, juneja2023assessing, zieringer2023algorithmic, haroonYouTubeGreatRadicalizer2022, ballardConspiracyBrokersUnderstanding2022}, including the spread of misinformation or how the filter bubbles tend to form on social media platforms and how the users can deal with them~\cite{tomlein_audit_2021, srba2023auditing, Hussein2020, yang2023bubbles, ledwich2022radical, kaiser2020birds, aridor2020deconstructing}. Furthermore, studies found that there is no difference in exposure to harmful content on TikTok between youth and adult accounts~\cite{xue2025towards}. However, in the case of disclosed and undisclosed commercial communication, our previous algorithmic audit revealed significant profiling aligned with minors' interests (5-8 times stronger than for adult formal advertising)~\cite{solarova2026dsasblindspotalgorithmic} .

A specific focus is dedicated to investigating and analysing how the recommender system behaves when it comes to polarising topics. Early works on this problem found that following the chain of recommendations on YouTube leads users more towards extreme content~\cite{Spinelli2020, Ribeiro2020}. However, recent studies have found opposing effects, where instead of pushing users into extremist bubbles, the recommender tends to lead users towards more mainstream or moderate content that is often more popular~\cite{ledwich2022radical}.

Many studies focus on political topics, particularly those related to elections. A large study of YouTube shorts focusing on the South China Sea dispute and the Taiwan presidential election found that the recommendations rapidly shift users toward entertainment or other non-political topics~\cite{cakmak2025investigating}. Studies focusing on the recommendations on TikTok and Twitter during the 2024 U.S. presidential election have found different findings~\cite{ye2025auditing, ibrahim2025tiktok}. One study found that for the new users, the algorithm tends to amplify right-leaning content, but for existing users, it tends to drift users towards their political views with minimal exposure to opposing perspectives (i.e., left-leaning users get more left-leaning videos and right-leaning users get more right-leaning)~\cite{ye2025auditing}. Another study found that the recommender on TikTok pushes users towards right-leaning content, even when the accounts are seeded with left-leaning videos~\cite{ibrahim2025tiktok}. In this case, the skew is primarily driven by negative partisanship content which criticises the opposing party, with left-leaning users being exposed more to the opposing views than right-leaning ones~\cite{ibrahim2025tiktok}. Similarly, a reverse engineering study found that TikTok tends to drift users towards far-right content, even without users actively seeking such content out, suggesting that the algorithm appears to actively contribute to radicalisation and polarisation~\cite {shin2024algorithms}. A study focused on the 2024 German election found that the recommender rewards divisiveness in political communication, where videos with negative emotions generate more engagement~\cite{solovev2025tiktok}. As such, studies also call for stronger regulation of short video format platforms, especially through better transparency~\cite{shin2024algorithms, soderlund2024regulating}.

In this work, we extend these existing works, primarily focused on politics, by comparing three types of personalisation drifts across multiple polarising topics and a neutral topic within the same experimental setup. Such analysis allows us to determine whether the drift is caused by the popularity of the topic and determine which topics the regulation should focus on more. In addition, we measure the personalisation-stance drift separately for the individual stances for the polarising topics (e.g., pro- vs. anti-vaccination), which allows us to draw conclusions about whether there is an overall drift towards one of the stances, whether the recommender reinforces the stance of the users or tries to also recommend opposing stances or more neutral topics. Furthermore, thanks to automatic LLM-driven user simulation, we continue to interact with relevant videos during the whole audit in order to further reinforce the user interests and achieve a more realistic behaviour. This allows us to explore the change in the recommended videos over time, while also leading to results more in line with typical users.

\section{Evaluating Personalisation Drift on TikTok}

\subsection{Audit Overview}

\textbf{Audit Scenarios.} To investigate and evaluate the personalisation drift in polarising topics on TikTok, we run a sockpuppeting algorithmic audit study using the web-based interface of the platform.  Each controlled user is assigned one of polarising or neutral topics and a supporting or opposing stance towards this topic (for a neutral topic, there is only one neutral stance).

The utilised audit scenarios consist of two main phases: 1) \textit{a seeding phase}; and 2) \textit{an interaction phase}. The purpose of the seeding phase is to build up and develop the user profile (i.e., interest in an assigned topic and stance towards it). It is conducted within the TikTok's search feature by searching for key phrases related to the assigned topic and watching videos that match the user's topic and stance. The purpose of the subsequent interaction phase is to measure three types of personalisation drifts we are interested in -- how the proportion of the videos with a specific topic and stance evolves over time. It is conducted within TikTok's \textit{For You} page by simulating user interactions with recommended videos. 

In both phases, simulated users should interact with searched or recommended content in line with their interests and stances. To this end, we created an automatic tool, which we call \textit{user interaction predictor}. Using a Large Language Model (LLM), it determines whether the video is relevant for the user and how the user should interact with it (i.e., given a video, what action should be taken).

For technical details describing the implementation of the auditing solution, please see Appendix B.

\textbf{Topics and Stances.} We focus on 4 topics with various level of polarisation: 1) \textit{flat earth}, with one side representing people believing that the earth is flat or that there are civilisations beyond the "ice wall of Antarctica" (support), and other side disputing these claims (oppose); 2) \textit{vaccination}, with one side claiming the vaccines are ineffective or can even cause harms (oppose), and other side opposing these claims and showing they are safe (support); 3) \textit{climate change}, with one side disputing that there is a climate change happening or what we are experiencing is a normal phenomenon or opposing the policies for fighting it (oppose), and other side disputing these claims and calling for better climate change awareness and fight against it (support); and 4) \textit{US politics}, with one side supporting right-wing politics and conservative voters (support), and other side opposing such politics and being oriented towards left-learning political views and liberal voters (oppose). Additionally, we use a neutral topic, \textit{cooking}, consisting of various recipes and other cooking-related content, as a baseline for comparing the drift effects. For this topic, we do not define different stances.

\textbf{Users.} Due to observed behaviour of TikTok and bots during the study (see the results Section~\ref{sec:rq1} for details), we specify 3 separate groups of users, each with slightly different \textit{seed} and \textit{interaction} phases. The first (\textit{neutral+polarising}) consists of 32 users (4 for each combination of topic and stance) who are interested in polarising and neutral topics in each phase. The second (\textit{polarising only}) consists of an additional 32 users (4 for topic+stance) that are only seeded with a polarising topic (representing maximum polarity), but interact with a neutral topic during the \textit{interaction} phase. Finally, the third (\textit{mixed polarity}) consists of 4 users only for the US politics topic that are seeded and interact with both stances at the same time (representing middle or neutral polarity). For all of these subsets, we used a 50:50 split for gender, and same age range (18-24). No user account in the study was banned by the platform, even though it employs strong detection capabilities.

The overall duration of the study is 16/10 days (1 day for the seeding phase + 15/9 days for the interaction phase for the first two and the third user group respectively), and it was run in September 2025 (the first two subsets) and January 2026 (the third subset) using proxy servers located in the USA. We have encountered more than 80000 unique videos, making manual annotation infeasible.

\subsection{User Interaction Predictor}

The user interaction predictor is designed to provide as realistic a user simulation as possible by automatically annotating the videos and determining which videos the user should interact with. Instead of relying on simple heuristics and static rules (which are common in the previous works, e.g., using sets of video hashtags \cite{boeker2022empirical}), in this study, we employ an LLM (specifically, \textit{GPT-4.1}) to accomplish this goal. The LLM is provided with the user characteristics (the topic of interest and stance), video URL and other video metadata obtained during the audit (title, description, author, video stickers). As many of the videos do not contain any usable author-provided description, or only very limited descriptions that do not by themselves accurately convey the topic or stance of the video, we download the audio track of the video using its URL, and then we use \textit{Whisper large-v3-turbo} model~\cite{radford2023robust} to get the transcript of the audio track.

The user characteristics, video metadata and voice transcript are used to construct a prompt for the LLM. The prompt we use is provided in Appendix A and was carefully created manually to achieve the highest possible performance. We specifically focus on prompt-engineering good practices by providing the LLM with all possible options, detailed description of the topics, their different stances and how they should be assigned. The prompt is dynamically constructed based on the topic, in order to perform only a three-way classification into topic of interest, neutral topic or unrelated topic, and using only the available metadata. The answer of the LLM is then parsed to determine whether the video is relevant. When the video is related to the topic and stance of interest for the user, the action returned by the user interaction predictor is to watch the video in full, like it and bookmark it. In any other case, the video is skipped. Furthermore, livestreams and any video longer than 5 minutes are automatically skipped.

We select the best-performing LLM by evaluating and comparing the performance of multiple LLMs (GPT-4o, GPT-4.1, LLaMA-3.1~\cite{grattafiori2024llama}, Gemma-2~\cite{gemma_2024} and Qwen-2.5~\cite{qwen2.5}) of different sizes and different prompt templates. To achieve this, we first constructed a set of simple queries for each topic and stance (e.g., "proof earth is flat" for the supporting stance of the flat-earth topic, or “debunking flat-earth theory” for the opposing stance). Using these queries, we manually collected and annotated videos belonging to each topic and stance of interest. We collected 50 videos for each out of 4 polarising topics (with an even 50:50 split between stances), 50 videos for the neutral topic, and an additional 100 videos not related to any of the topics in the study. Using these videos, we constructed an evaluation dataset comprising 350 videos and utilised it to evaluate the LLM with the constructed prompt. The best performing model (GPT-4.1) achieves topic classification accuracy of $98\%$, $95\%$, $98\%$ and $96.5\%$ for flat earth, vaccines, climate change, and US politics topics, respectively. The stance classification is evaluated only on 50 videos belonging to the topic, as the stance for other videos is not relevant. In this case, we observe the accuracy to be $100\%$, $90\%$, $98\%$ and $94\%$ for flat earth, vaccines, climate change, and US politics topics, respectively. After deeper analysis, the errors are only due to false positives (for topic classifications) and mixed stance videos, where even human annotators had a lower agreement (for stance classification). Based on these high scores, we are sufficiently certain our user interaction predictor is capable of performing its task. 

We also evaluated the performance of different versions of the Whisper model (small, base, medium, large, turbo)~\cite{radford2023robust} on the same set of 350 videos. We found that adding audio transcripts to GPT-4.1 prompts, on average, increases the topic prediction by $2\%$ and stance prediction by $6\%$. We finally chose to use \textit{large-v3-turbo} model because the predictor accuracy using this version was only marginally smaller, but the model was substantially faster, enabling us more efficient usage of available computational resources as well as to simulate an immediate user reaction after the start of video playback, which is crucial in short-form TikTok videos.

\subsection{Audit Phases}
\label{sec:audit_phases}

\begin{figure}[t]
    \centering
    \includegraphics[width=1\linewidth]{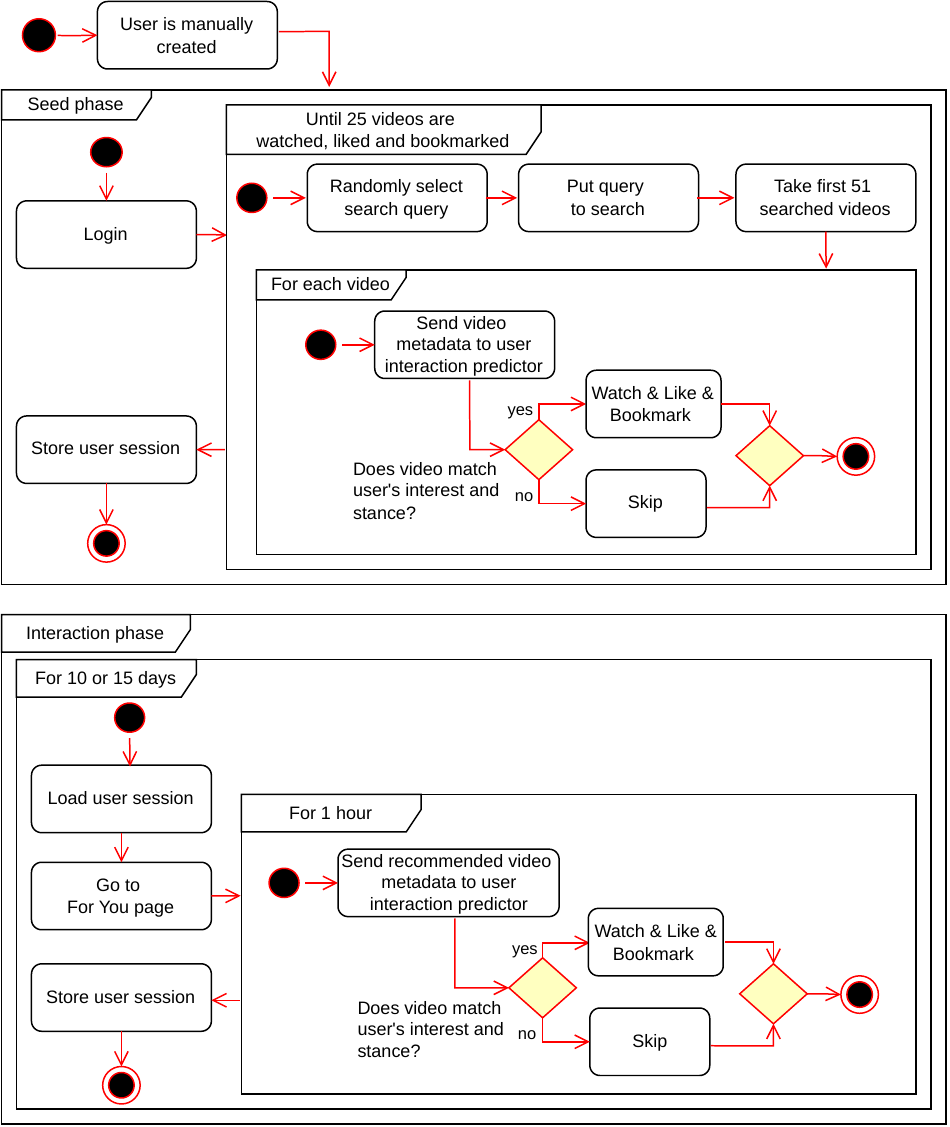}
    \caption{Visualisation of the audit phases.}
    \label{fig:process-diagram}
\end{figure}

In this section, we provide a detailed description of the individual audit phases and how they are performed. The overview of these phases is illustrated in Figure~\ref{fig:process-diagram}.

Before starting the audit itself, we prepared a list of queries for each topic and its stance in the study (the list and how it was created is described in Appendix A). Afterwards, we created the individual user accounts with given characteristics and perform a first login, extracting cookies for easier access later.

\textbf{Phase 1: Seeding the users.}
The objective of this phase is to mimic users' interests and history so the TikTok recommender system can quickly pick up on the user preferences and serve related content. To achieve this, each user watches a predetermined number of videos (25) belonging to their interest. First, we randomly select a query from the list of queries corresponding to the topic and stance of the user and use it to search for videos on TikTok (using the search functionality). From the returned videos, we take the first 51 and pass them to the \textit{user interaction predictor} to determine whether they are of interest to the user. The videos matching the interest are clicked on, watched in full, liked and bookmarked. For the remaining videos (not of interest), there is no interaction. These actions are repeated until a sufficient number of videos of interest are watched, using multiple randomly sampled queries if necessary (e.g., when one search query does not contain a sufficient number of videos of interest). In preliminary experiments, we have confirmed that 25 videos watched over 1 day is enough to sufficiently build the users' profile.

For the first user group (\textit{neutral+polarising}), we start the seed phase by watching videos from the neutral topic (cooking) until 25 videos are watched, followed by the polarising topic of interest, again until 25 such videos are watched (as such, watching a total of 50 videos in the seed phase). In addition, after running the \textit{interaction phase} for 3 days, we repeat the seeding with an additional 25 videos from the polarising topic. For the second group (\textit{polarising only}), we use only the polarising topics for seeding, watching 25 videos. Finally, for the third group (\textit{mixed polarity}), the users are seeded with 25 videos from both stances (50 overall) in no particular order. In all cases, after running the seed phase, we implement a wait time of 1 day to give the recommender system time to take the interests into consideration.

\textbf{Phase 2: Interacting with For You Page. }
After the user is seeded and the 1-day wait time has passed, we initiated the interaction audit phase, observing how the videos recommended by TikTok on the \textit{For You} page change over a 15-day period. Each day, the user logs in (reusing cookies from the first login), opens the \textit{For You} page and interacts with the recommended videos for up to approx. 1 hour (the average time a real user spends on the TikTok platform per day). Each encountered video is sent to the \textit{user interaction predictor}. Only the videos corresponding to the topic and stance of the user (in case of the third user group, this includes both stances), or the neutral topic (only for first two use groups) are interacted with, watched in full, liked and bookmarked. All remaining videos are skipped after a short delay (1-2 seconds), which simulates the real user and the time it takes to decide whether the video is of interest. This way, we are not skipping immediately, which could be flagged by the platform as an artificial behaviour, but also not watching for a long period of time, which would give an implicit feedback signal to the recommender system. In addition, this behaviour further reinforces the user interests. For each video encountered, we save the URL and metadata (title, description) and the prediction from the \textit{user interaction predictor}.

\subsection{Evaluation methodology}

To evaluate the drift, we examine two aspects. First, we track the overall number of videos that belong to the topic of interest of a given user. Second, we track the number of videos with both stances towards the topic of interest.

As each user may observe a different number of videos during the day, we first collate the data from individual runs of the user as we are interested in the continuous drift over the whole time of the study. This allows us to determine whether there are pronounced periods of exploration (where the recommender tends to provide recommendations that have different topic/stance from users exhibited preferences) or exploitation (where the recommender tends to provide recommendations that have topic/stance aligned with users exhibited preferences). To perform this preprocessing, we first create intervals of 30 minutes (i.e., bins), essentially splitting a 60-minute-long daily session into two parts. Afterwards, we partition the user interactions into these bins and aggregate over them. If any video is encountered after the predefined 60 minutes in the day, it is placed into the second bin. Then, for each bin, we count the number of overall videos the user is recommended, the number of videos belonging to the topic of interest (and also to the neutral topic when evaluating RQ1) and from them also the split between stances. When aggregating over multiple users, we take the ones with the same interests and add together the video counts (e.g., if we compare polarising topic, we take all users from the topic regardless what stance they were seeded with and count their videos in the given bin). Using these counts we then calculate either the ratio of videos belonging to the topic of interest from all videos that were recommended, or the ratio between different stances of the given topic.

For the \textit{preference-aligned} drift we calculate the change in ratio between the personalised video recommendations (topic of interest videos plus neutral topic videos) and random videos across each bin. For the \textit{polarisation-topic drift}, we calculate the change in ratio between the number of recommended videos from polarising topic and neutral topic across each bin. In this case, the value of $1$ represents that all videos are from the polarising topic, while $-1$ represents that all videos are from the neutral topic. For the \textit{polarisation-stance} drift, we calculate the change in ratio between the number of recommended videos with support stance and oppose stance across each bin. In this case, the value of $1$ represents all videos are from support stance, while $-1$ represents all videos are from oppose stance. We visualise the drift by fitting a regression model on the observed results. Finally, we also use the Mann-Whitney U test to determine the significance of the differences. As with all the phases of the study, the topic and stance are assigned by the \textit{user interaction predictor}.

\section{RQ1: Comparing Drift Between Neutral and Polarising Topics}
\label{sec:rq1}

\begin{figure}[t]
     \centering
     \begin{subfigure}[t]{0.49\linewidth}
        \centering
        \includegraphics[width=1\linewidth]{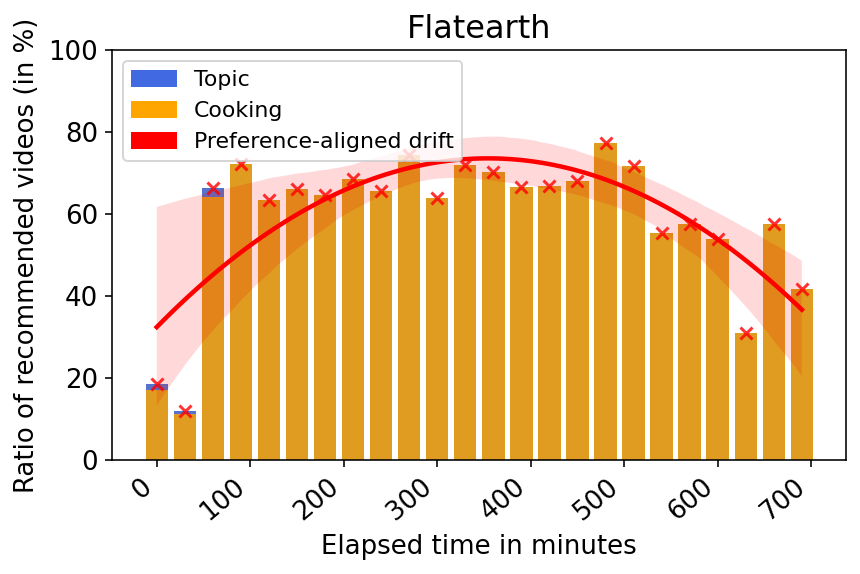}
    \end{subfigure}
    \begin{subfigure}[t]{0.49\linewidth}
        \centering
        \includegraphics[width=1\linewidth]{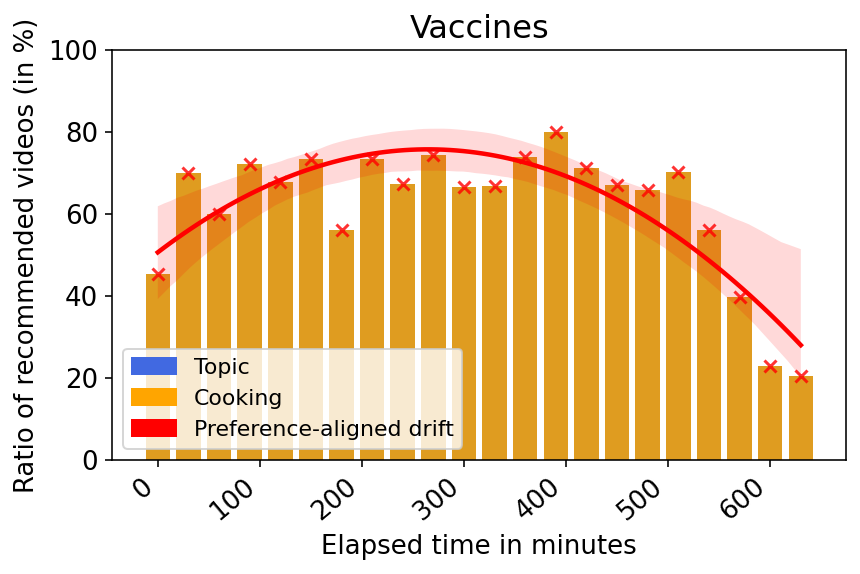}
    \end{subfigure}
    \hfill
    
    \begin{subfigure}[t]{0.49\linewidth}
        \centering
        \includegraphics[width=1\linewidth]{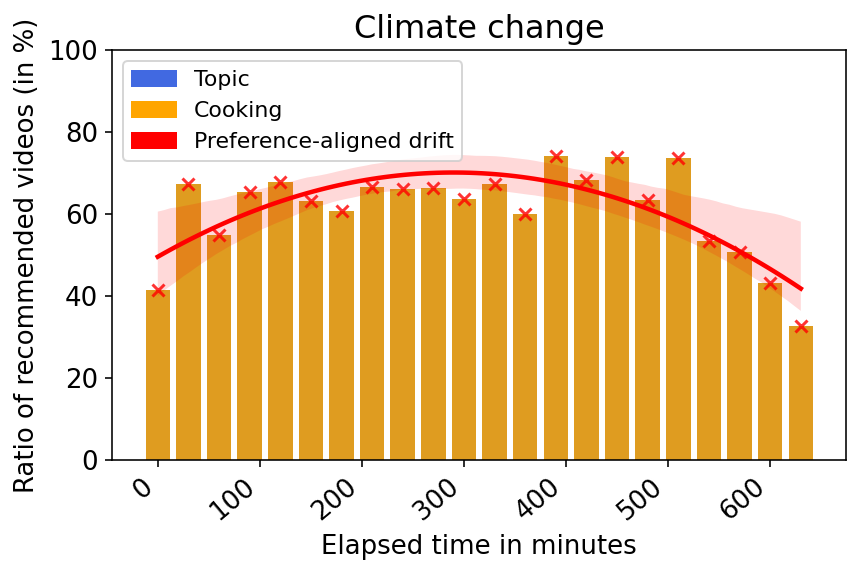}
    \end{subfigure}
    \begin{subfigure}[t]{0.49\linewidth}
        \centering
        \includegraphics[width=1\linewidth]{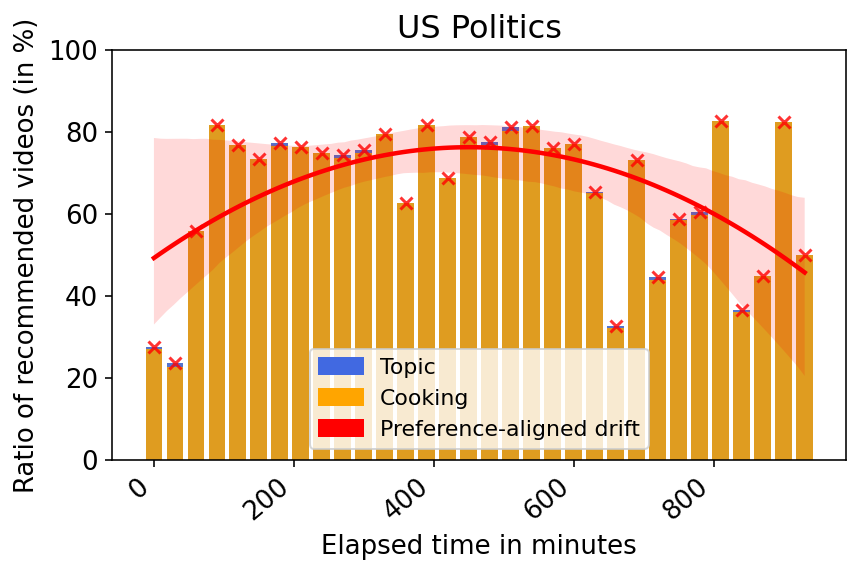}
    \end{subfigure}
    
    \caption{Ratio of videos for the \textit{polarising} and \textit{cooking} topics over time for users seeded with both topics (\textit{neutral+polarising}). We can see that the neutral cooking topic completely dominates the personalisation across all polarisation topics.}
    \label{fig:cooking-domination}
\end{figure}

In this section, we are mainly interested in the \textit{preference-aligned} and \textit{polarisation-topic} drift, utilising the first user group (\textit{neutral+ polarising} -- seeded with both neutral and polarising topics). From the overall encountered videos, we count how many belong to the topic of interest, the neutral topic and how many are unrelated, and report this over time. The results, along with the drift, are shown in Figure~\ref{fig:cooking-domination} for each polarising topic and aggregated over all users with the given interest (total number of 32 users, 8 for each topic).

As can be expected, we observe a strong \textit{preference-aligned drift}, where the number of videos of interest quickly increases to around $70\%$. At the same time, we can observe that even though we use the same number of videos for seeding the neutral and polarising topic (starting with the neutral one), the videos from the neutral topic completely dominate the recommendation list, with almost no videos from the polarising topics. Only US politics show a small number of polarising topic-related videos. In this case, we also run a repeated seed phase after the third day, but only for the polarising topic, in order to promote it for the recommender system. However, this has only a minimal effect on the observed distribution of recommended videos, as the cooking videos continue to dominate. As such, we can conclude that when combining the polarisation and neutral topics as interests for the user, the TikTok recommender system rapidly shifts towards more neutral and safe videos for the user. Such behaviour, however, may also be caused by the higher popularity and prevalence of the neutral topic.

\begin{figure*}[ht]
    \centering
    \hspace{.1\linewidth}
    \begin{subfigure}[t]{0.33\linewidth}
        \centering
        \includegraphics[width=.95\linewidth]{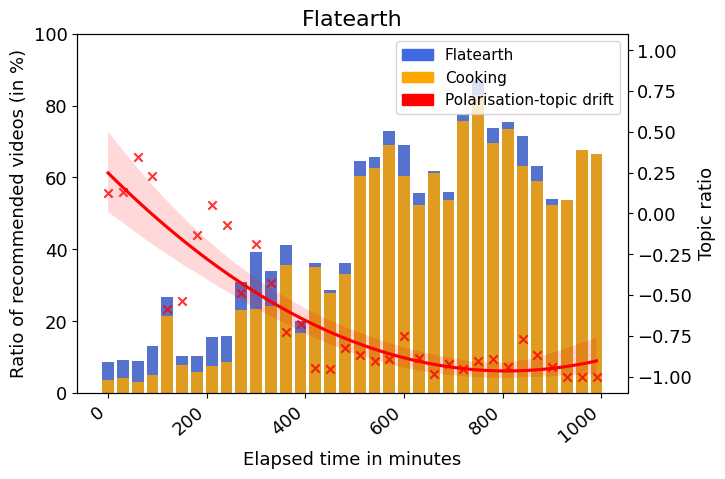}
    \end{subfigure}
    \hfill
    \begin{subfigure}[t]{0.33\linewidth}
        \centering
        \includegraphics[width=.95\linewidth]{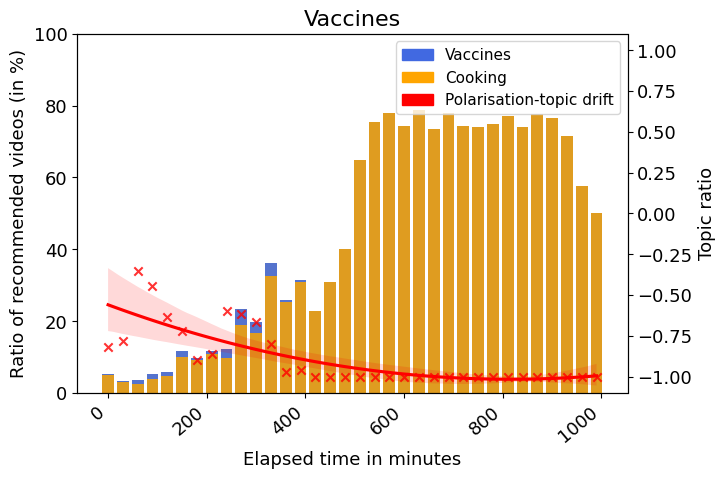}
    \end{subfigure}
    \hspace{.1\linewidth}
    \hfill
    
    \hspace{.1\linewidth}
    \begin{subfigure}[t]{0.33\linewidth}
        \centering
        \includegraphics[width=.95\linewidth]{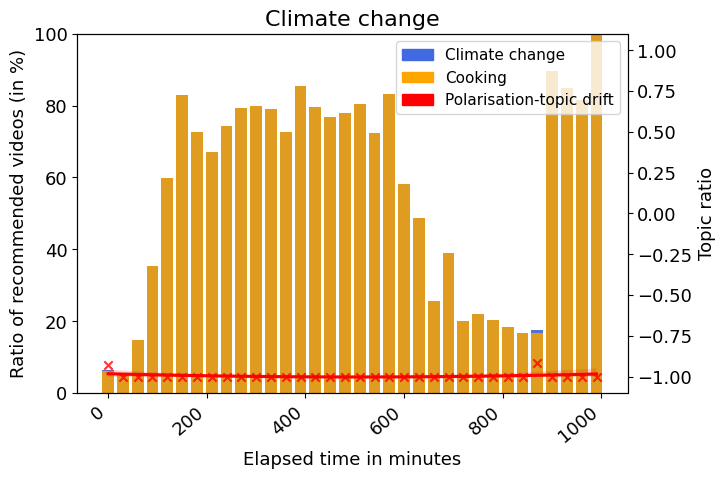}
    \end{subfigure}
    \hfill
    \begin{subfigure}[t]{0.33\linewidth}
        \centering
        \includegraphics[width=.95\linewidth]{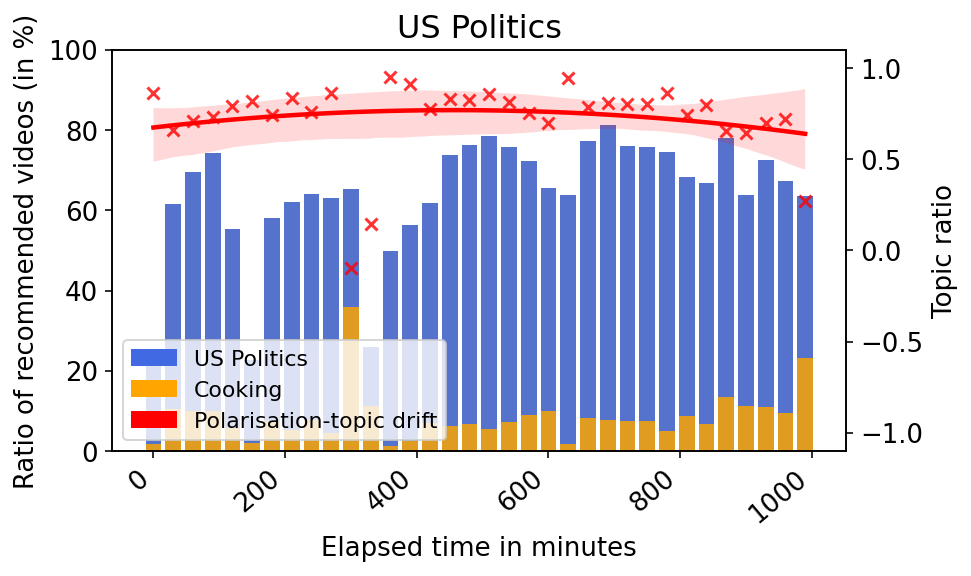}
    \end{subfigure}
    \hspace{.1\linewidth}
    
    \caption{Ratio of videos for the \textit{polarising} and \textit{cooking} topics over time for users seeded with the polarising topics only (\textit{polarising only}). We observe a \textit{strong neutralising polarisation-topic drift} for climate change and vaccines, while US politics show an equilibrium behaviour with a large ratio of political videos.}
    \label{fig:cooking-vs-polarising}
\end{figure*}

To provide further analysis on how the polarisation topics are treated, we use the second user group (\textit{polarising only} -- seeded only with polarising topics, while cooking videos are still watched during the interaction phase when encountered). The results of this analysis, along with the \textit{polarisation-topic} drift, are shown in Figure~\ref{fig:cooking-vs-polarising} for individual topics and aggregated over all the users with the given interest (total number of 32 users, 8 for each topic).

In this case, we observe a more gradual \textit{preference-aligned} drift. The only exception is US politics, where the switch to observing majority of topic-related videos is much faster (immediately from the start). For the \textit{polarisation-topic} drift, we observe significant differences between topics. For the misinformation-related topics, we see more of a \textit{neutralising effect}. At the start, the majority of encountered videos are unrelated, with only a few cooking or topic videos. As time goes on, the recommender picks up on the occasional interest in cooking videos, which start dominating the feed towards the end, completely eliminating videos from the polarising topic. As such, we can detect signs of topic suppression over time. This behaviour is strongest for climate change (only $6$ videos from the polarising topic are encountered), followed by vaccines (where we observe a few polarising videos at the start, up to a total number of $126$; with cooking topic representing more than $50\%$ videos only towards the end). For flatearth, the recommender picks up on the interest in the polarising topic and starts recommending an increasing number of videos (overall $735$ videos watched). However, even here, the neutral topic starts dominating after 5 days.

On the other hand, for US politics, we observe completely different results, with an overall number of $5117$ videos. After the first day, the number of videos related to the topics of interest grows rapidly, until reaching approximately $60-80\%$ of all videos, showing a very strong \textit{preference-aligned drift}. At the same time, although the recommender picks up on the neutral cooking topic, recommending approximately $25\%$ of the overall number of videos, it never outgrows the polarising topic. As such, we do not observe any \textit{polarisation-topic drift}. Instead, the systems seems to keep an equilibrium state, with the ratio of both the US politics and cooking videos remaining stable over the whole time. In addition, we can observe exploration phases towards the end of day 3 and day 6 of the audit, where the relevant videos represent only about $20\%$ of the videos (corresponding to the start of the audit which is also defined by strong exploration). Besides these two exploration phases, the TikTok recommender seems to prefer exploitation, providing only about $10-20\%$ of videos beyond users' interests. Similar behaviour, although with a weaker exploitation phase, can be observed for the flatearth topic as well. For the remaining topics, the exploitation phase (with cooking videos) ends up at $80\%$ videos as well (which seems to be the top thresholds for a number of relevant videos).

Overall, we can conclude that TikTok recommender treats the polarising topics in a substantially different manner, either due to their character (more misinformation-based vs. only polarising with no "immediate significant harms") or their popularity and representation on the platform (e.g., political videos being more popular than vaccination or climate change). For the more misinformation-based topics, namely climate change, vaccines and flatearth, we observe neutralising \textit{polarisation-topic drift} and weaker \textit{preference-aligned drift}, with no polarising videos observed for climate change. Finally, for the US politics, we observe no \textit{polarisation-topic drift} but instead an equilibrium (same large number of polarising videos over the whole time) and a strong \textit{preference-aligned drift}. The suppression of the more misinformation-based topics and dominance of the neutral topic cannot be explained solely through the higher popularity and prevalence of videos from neutral topic. Even though both US politics and vaccines are very popular topics, based on the popularity of highest occurring hashtags (around 20-30 million hashtag uses), the behaviour of the recommendation algorithm is quite different for these two topics (see Appendix D for the hashtag popularity analysis).

\section{RQ2: Comparing Drift Between Stances of Individual Topics}
\label{sec:rq2}

\begin{figure*}
    \centering
    \centering
    \begin{subfigure}[t]{0.33\linewidth}
        \centering
        \includegraphics[width=.95\linewidth]{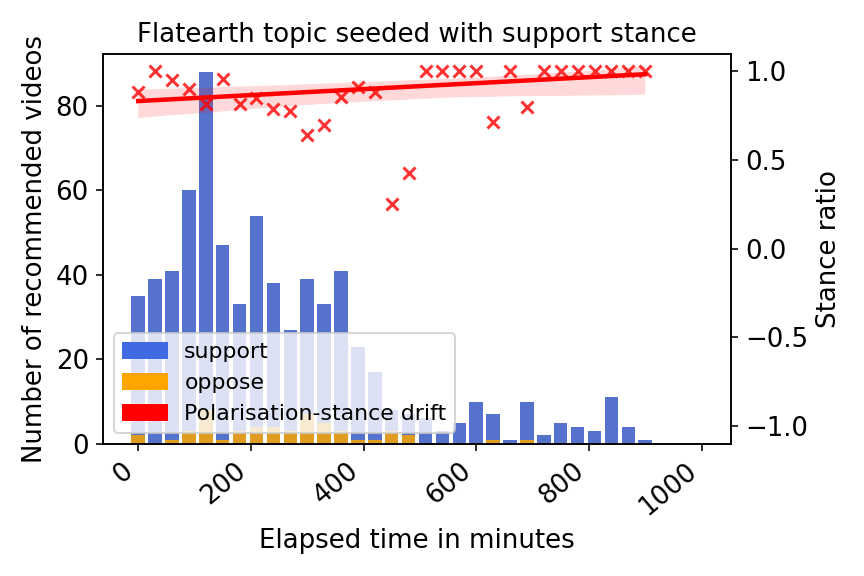}
    \end{subfigure}
    \hfill
    \begin{subfigure}[t]{0.33\linewidth}
        \centering
        \includegraphics[width=.95\linewidth]{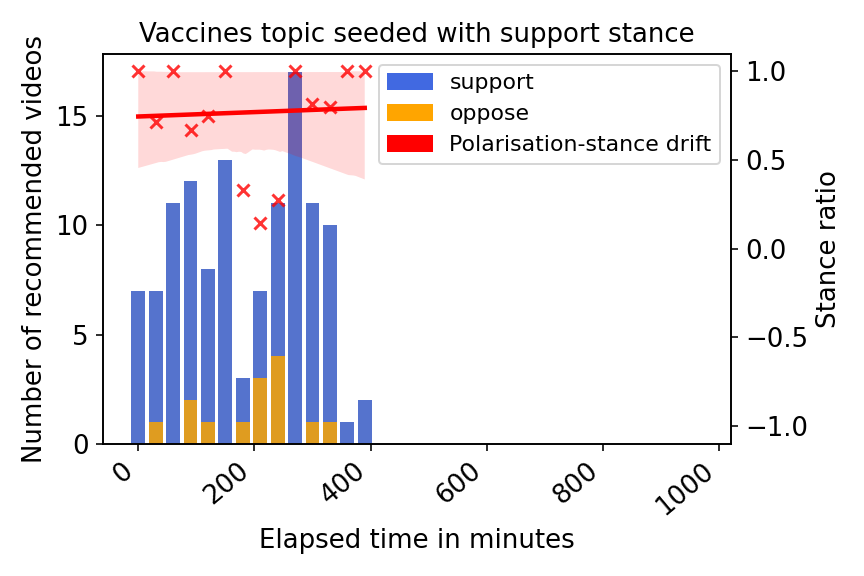}
    \end{subfigure}
    \hfill
    \begin{subfigure}[t]{0.33\linewidth}
        \centering
        \includegraphics[width=.95\linewidth]{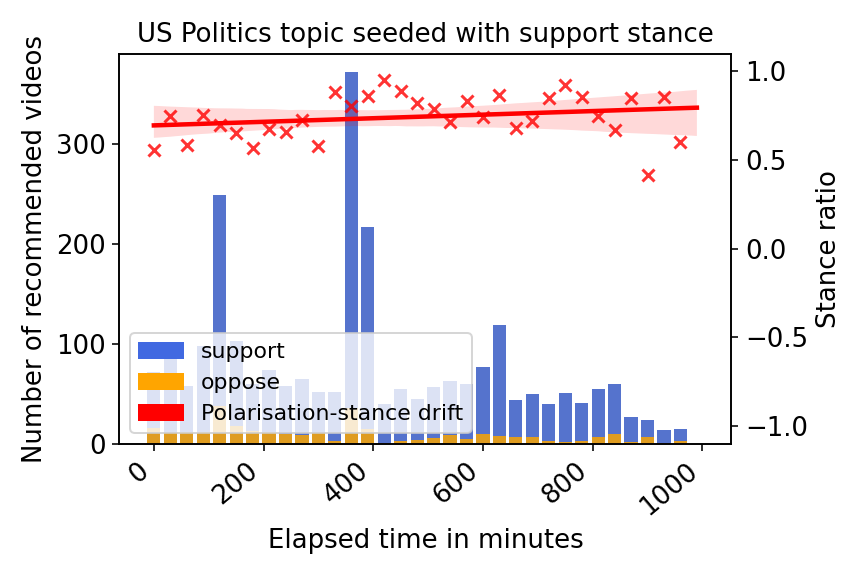}
    \end{subfigure}
    \hfill
    \begin{subfigure}[t]{0.33\linewidth}
        \centering
        \includegraphics[width=.95\linewidth]{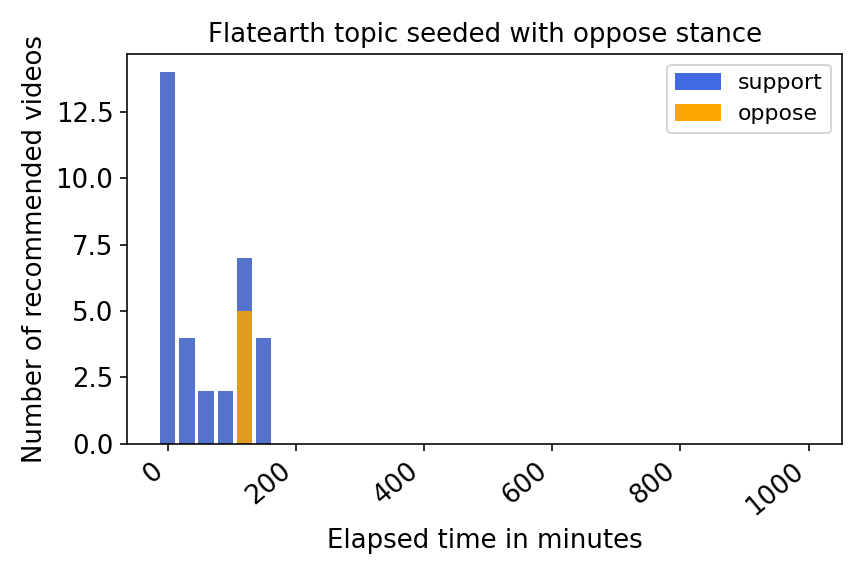}
    \end{subfigure}
    \begin{subfigure}[t]{0.33\linewidth}
        \centering
        \includegraphics[width=.95\linewidth]{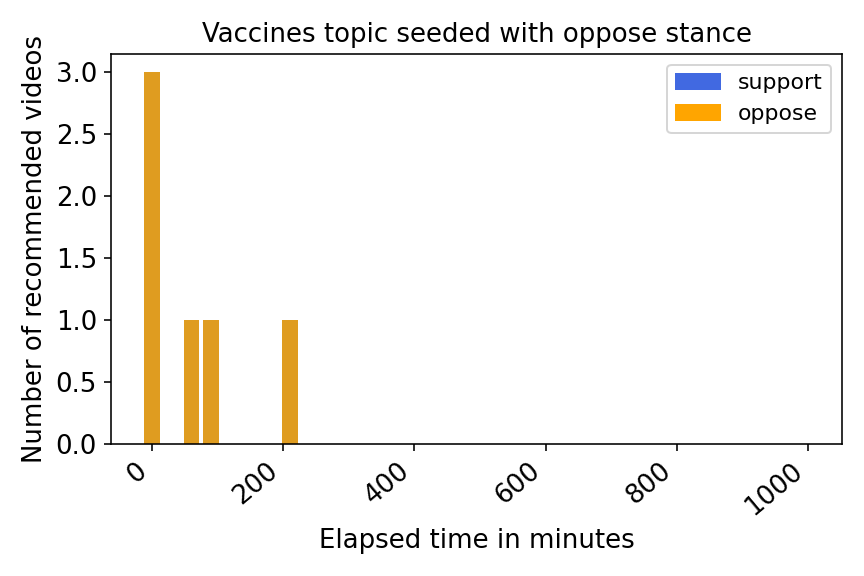}
    \end{subfigure}
    \hfill
    \begin{subfigure}[t]{0.33\linewidth}
        \centering
        \includegraphics[width=.95\linewidth]{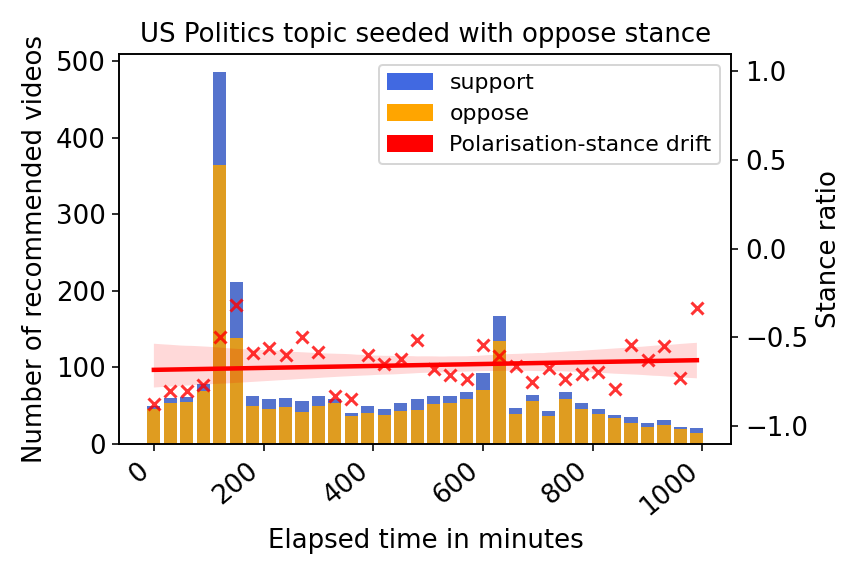}
    \end{subfigure}
    \caption{Average number of videos with specific stance across topics. We observe an equilibrium behaviour for \textit{polarising-stance drift} with similar number of videos from the seeded stance over time and only few videos from opposing stance.}
    \label{fig:topics-and-stances}
\end{figure*}

In this section, we examine the \textit{polarisation-stance drift} for the individual polarising topics and their corresponding stances using the second user group (\textit{polarising only}). We are mainly interested in whether we can observe a drift towards a specific stance regardless of the seed (i.e., polarising towards \textit{support} or \textit{oppose}) or the recommender tends to treat each stance the same way, and either push users towards their stance of interest (i.e., \textit{polarising effect}), pushes users to the opposing views (i.e., \textit{negating effect}), or tries to include both sides of the issue (i.e., \textit{balancing effect}). Finally, we also want to determine whether all topics are treated equally. The results of this analysis, along with the \textit{polarisation-stance drift}, are presented in Figure~\ref{fig:topics-and-stances}. The results for \textit{climate change} topic are not included due to the strong neutralising effect demonstrated in the previous section.

Overall, we can observe an equilibrium behaviour, with small differences between topics. For the flatearth and vaccines topic seeded with opposing stances (i.e., videos that debunk flatearth and videos that spread vaccine misinformation), we observe a minimal number of videos that would even belong to the topic (on average $8$ for flatearth and $2$ for vaccines over 15 day period). As such, we cannot draw any conclusions for these cases. On the other hand, for the support stance in these topics, we observe a significantly larger number of topically-related videos, with a clear dominance of the supporting stance for flatearth (the p-values of 9.2e-11), but no significant difference for vaccines (p-value of 0.14). In both cases, we do not observe any specific drift. Interestingly, we observe an overall larger number of videos with opposing stances in this case than when the opposing stance was directly used for seeding ($53$ for flatearth and $14$ for vaccines).

In the case of US politics, the different stances appear to be treated equally, with no significant difference between number of stance-related videos (p-value of 0.939). In both cases, we observe a significantly large number of videos that correspond to the seeded stance (p-values of 2.331e-13 for support and 9.800e-36 for oppose), with occasional videos from the inverse stance (e.g., opposing videos for users seeded with a support stance). In addition, for both stances we observe significant spikes at the same time intervals, where the number of videos belonging to the topic is as high as $400-500$. Such spikes appear more often for the users seeded with the support stances, but it may not represent any tendency of the recommender system. However, there is no observed \textit{polarisation-stance drift} towards any of the stances.

\begin{figure*}
    \centering
    \hspace{.1\linewidth}
    \begin{subfigure}[t]{0.3\linewidth}
        \centering
        \includegraphics[width=.95\linewidth]{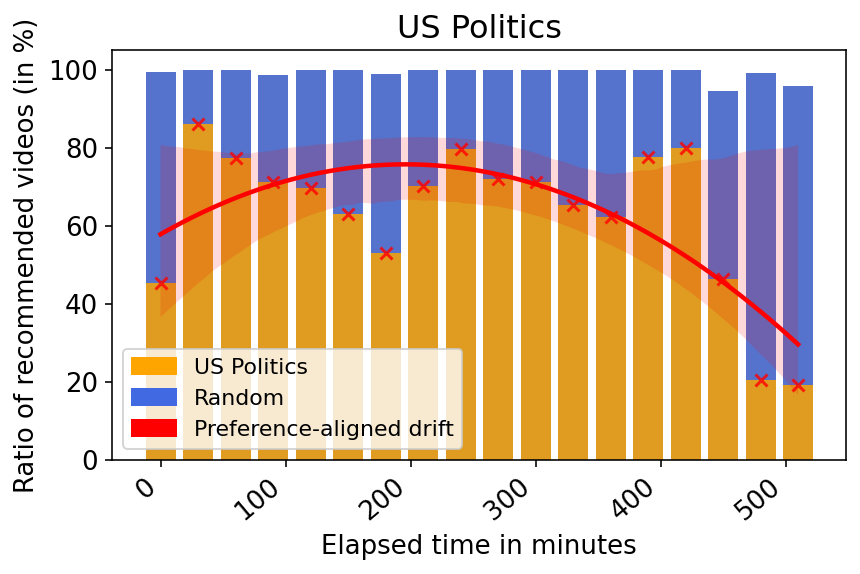}
    \end{subfigure}
    \hfill
    \begin{subfigure}[t]{0.3\linewidth}
        \centering
        \includegraphics[width=.95\linewidth]{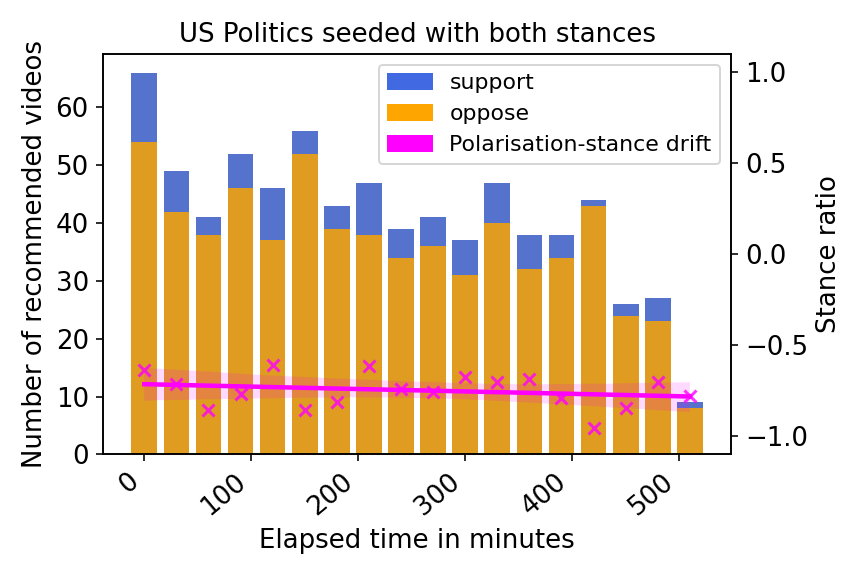}
    \end{subfigure}
    \hspace{.1\linewidth}
    \caption{\textit{Preference-aligned} (left) and \textit{polarisation-stance} (right) drift for users seeded equally with both stances for the US politics topic (\textit{mixed polarity}). We observe a stronger tendency of the system to recommend videos from the oppose stance.}
    \label{fig:newest-results-us-politics}
\end{figure*}

To provide more detailed analysis and to investigate whether there is a stronger tendency to recommend a specific stance (which the second user group cannot be used for due to how it was seeded), we use the final third group seeded in an equal manner (\textit{mixed polarity}). The analysis is done only for the US politics topic, as it is the single topic where we observe a sufficient number of polarising videos. The results, along with the \textit{preference-aligned} and \textit{polarisation-stance} drift, are presented in Figure~\ref{fig:newest-results-us-politics}.

For the \textit{preference-aligned drift}, we observe similar results as before, where the system recommends majority of the videos of interest, but slowly moves towards exploration. For the \textit{polarisation-stance drift}, we see a significant preference (p-value of $2.113e-22$) for the oppose stance. At the same time, we observe small drift towards oppose videos (drift starting around $-0.7$ and ending at $-0.8$) although both polarities are watched. As such, this indicates that in the US politics topic, TikTok recommender system has a \textit{polarisation-stance drift} towards the oppose stance. However, to a certain extent, this may also be an effect of video popularity and recency, due to the events happening in the US during the time of our audit. To determine the long-term drift, the audit would need to be run for a significant longer time (e.g., months) to capture these nuances.

Overall, we can conclude that there is no specific \textit{polarisation-stance drift} when users are interested only in a single stance (with only occasional videos from the inverse stance). However, when users are interested equally in both stances, there is a small \textit{polarisation-stance drift} towards oppose videos for US Politics topic. For the misinformation-themed topics, there seems to be a tendency to recommend a larger number of videos supporting the topic (i.e., spreading flat-earthist viewpoints, or supporting vaccination), which can at least partially stem from the larger presence and popularity of such videos on the platform.

\section{Ethical Considerations and Limitations}

The research presented in this paper was done as a part of the research project, which obtained approval from the organizational Ethics Committee (decision as of December 17, 2024). Researchers and research engineers conducting this auditing study also participated in four ethics assessment workshops together with ethics and legal experts, where relevant ethical and legal challenges have been identified and appropriate mitigations proposed. More specifically as a result of ethics assessment workshops we identified 36 potential ethical risks relevant to algorithmic auditing. We evaluated each risk with respect to exposure, likelihood and risk severity and identified potential countermeasures that could be used to mitigate these risks. In the rest of this section, we present the most important risks (with a high likelihood and severity) relevant to this study.  

Among the most important identified risks was processing of personal and sensitive data. To mitigate this risk, we limited the scope of collected data to not include personal information such as the name of creator. When some personal data is inevitably collected, processed, or stored, we ensure compliance with the GDPR, internal Data Research Policy and other relevant regulations. 

Another important risk we identified is the risk of breaching of the terms of service of a TikTok platform, which could be a result of creating automated bots and using them for data collection. 
However, this breach of ToS is permitted by Article 40 (12) of the EU Act on Digital Services (DSA) if the research concerns systemic risks. Understanding the polarisation drift can be viewed as a systemic risk, as foreseen by Recital 83 of the DSA.

The interaction of the bots with the content on the platform may impact the platform and society (e.g., increasing the view or like count). We mitigate this risk by minimizing the number of bots that we run and collecting only publicly available metadata. 

Finally, the \textit{user interaction predictor}, which we use for annotation purposes in evaluation, is based on large language models, and so we may observe potentially biased and incorrect findings due to the mistakes made by it. We address this risk by systematically evaluating and reporting the potential error rate. Furthermore, to accomplish this, we need to perform human annotation. It is done solely by the authors of the study, following recommendations from ethics experts in order to minimise possible negative consequences and ensure well-being. Finally, to minimise any potential legal and ethical issues, we directly involve legal and ethics experts.

\textbf{Limitations.} The audit is run with newly created bot accounts, which may be treated differently by the platform (i.e., providing a stronger exploration element over exploitation) and thus bias the observed results. In addition, we focus only on the USA and as such, the results may not generalise to other countries. Furthermore, the results, as with all audits of this type, are tied to the current time period due to the popularity of the content. We plan to address this limitation as future work by providing longitudinal and multi-platform large-scale audits.  The videos for evaluating the user interaction predictor were collected manually through the search functionality and may be biased (e.g., it could have happened that videos that are not as straightforward were skipped or there is not enough videos for the topics to provide statistically significant results), leading to different real performance of the LLM.  Finally, we use only a single neutral topic (which we found during the study to be highly popular), which may not sufficiently represent a complex behaviour of real-world users. Finally, we also explore only strong feedback signal through watch, like and bookmark actions at the same time. Evaluating other combination of feedback signals as part of study of this kind is relegated as future work.

\section{Conclusion}

In this work, we systematically assess three types of personalisation drift (preference-aligned, polarisation-topic and polarisation-stance) in polarising topics caused by TikTok's recommender systems. To this end, we conducted an algorithmic audit in which we simulated 68 users with an interest in 4 polarising topics, with a supporting and opposing stance towards such topics. During a time period of up to 16 days, we were observing how personalisation within the recommender systems develops.   

We can conclude that the behaviour of the TikTok recommender system is significantly topic-dependent. First, we observe a strong personalisation towards the topics of interest. Second, we observe a neutralising polarisation-topic drift with various strengths, which is evident especially for the more misinformation-themed topics (climate change, vaccination and, to some extent, also flat earth), where we  observe the neutral topic to completely dominate the feed. For the US politics topic, we observe a tendency to show a majority of videos belonging to the political topic with no specific drift (i.e., keeping an equilibrium). Third, when seeded with a single stance, we observe the tendency for equilibrium behaviour as well, providing only videos from the specific stance. At the same time, we observe more videos for the support stance for flatearth and vaccines topics, while the same amount of videos for the US politics topic. On the other hand, when seeded with both stances equally, we observe a strong tendency and drift towards the oppose stance in the US politics topic (which may be explained by the events happening in the country at the time the audit was conducted).

These findings provide a valuable insight into the internal workings of TikTok recommender systems. They reveal how the complex and dynamic interplay between user-generated content, AI-driven algorithms, and user interactions can substantially shape the content that users are exposed to on the platform.

In future work, we aim to extend this approach to additional platforms, to increase the authenticity of user behaviour, as well as to audit additional phenomena.

\begin{acks}
This work was partially funded by the EU NextGenerationEU through the Recovery and Resilience Plan for Slovakia under the project \textit{AI-Auditology}; and by the European Union NextGenerationEU through the Recovery and Resilience Plan for Slovakia under the project No. 09I03-03-V04-00336.
\end{acks}

\bibliographystyle{ACM-Reference-Format}
\bibliography{references}

\appendix

\section{Additional Methodology Details}
\label{app:additional_details}

In this section, we provide supplementary information regarding the algorithmic audit methodology. The queries we use for seeding the individual topics and stances (and used in collection of the videos to test the user interaction predictor) are presented Table~\ref{tab:queries}. To create these lists, we first brainstorm simple common queries for each topic and stance (e.g., "earth is flat" for flatearth support topic). Afterwards, we further extend this set by prompting a large language model (in our case, GPT-4.1) to generate possible queries, given the ones we already have (giving them to the LLM as examples in the prompt). We manually check each of the queries on TikTok search to determine whether they are allowed (i.e., return any videos at all or there is message from the platform that the topic is not available because of its nature, as is often common with climate change) and whether the returned videos belong to the topic.

\begin{table*}[!tbh]
\centering
\footnotesize
\caption{Queries used for seeding the individual topics and stances}
\label{tab:queries}
\begin{tabularx}{\textwidth}{@{}lX@{}}
\toprule
\textbf{Topic+Stance} & \textbf{Queries} \\ \midrule

flatearth support & “proof earth is flat”; “flat earth facts they don’t want you to know”; “NASA lies flat earth truth”; “real photos of flat earth”; “why the globe is a hoax”; “best flat earth TikTok accounts”; “flat earth community on TikTok”; “flat earth videos debunking globe”; “viral flat earth TikToks”; “flat earth experiments”; “flat earth explained simple”; “how flat earth works”; “flat earth map vs globe map”; “how gravity works on flat earth”; “flight paths that prove flat earth”; “government hiding flat earth”; “why science is wrong about earth”; “flat earth truth banned videos”; “schools lie about earth shape”; “hidden flat earth documents”; “sun and moon on flat earth model”; “edge of the world Antarctica wall”; “flat earth dome theory animation”; “flat earth experiments you can try”; “stars and planets on flat earth”; “uncovering truth about flat earth”; “seek the truth flat earth”; “flat earth proof firmament beyond”
 \\ 
flatearth oppose & “debunking flat earth theory”; “science proves earth is round”; “flat earth vs globe explained”; “how we know earth is a sphere”; “physics that destroys flat earth claims”; “funniest flat earth TikToks”; “flat earther gets owned”; “flat earth logic fails”; “watch flat earther get roasted”; “globehead reacts to flat earth TikToks”; “real earth curvature footage”; “globe model experiments”; “proof of gravity and round earth”; “responding to flat earth claims”; “scientist reacts to flat earth TikTok”; “teacher explains why earth isn’t flat”; “pilot explains earth curvature”; “astronauts prove earth is round”; “why do people believe in flat earth?”; “how flat earth conspiracy spreads”; “flat earth is just a cult”; “watch me argue with a flat earther”; “disproving flat earth theories” \\ \midrule

vaccines support & “how vaccines work explained”; “why vaccines are safe and effective”; “vaccine science made simple”; “covid vaccine saves lives”; “what anti-vaxxers get wrong”; “debunking anti-vax claims”; “doctor reacts to anti-vax TikTok”; “myth vs fact: covid vaccine”; “watch this anti-vaxxer get fact checked”; “explaining vaccine side effects”; “vaccines and herd immunity”; “how vaccines stopped pandemics”; “why vaccine mandates make sense”; “nurse explains vaccine side effects”; “why I got vaccinated as a doctor”; “covid ICU stories from unvaccinated”; “frontline workers vs anti-vaxxers”; “calling out anti-vax influencers”; “fighting vaccine misinformation”; “spreading facts not fear”; “covid vaccine is not a conspiracy”; “protect the vulnerable get vaccinated”; “anti-vaxx logic be like…”; “watch this anti-vaxxer get owned”; “funniest anti-vax TikToks”; “satire: anti-vax commercial parody”; “this anti-vax argument makes no sense”; “vaccine privilege in the west”; “global vaccination efforts 2025”; “why vaccines matter worldwide”; “covid isn’t over for everyone”; “vaccine equity explained” \\
vaccines oppose & “why vaccines are dangerous”; “do not get vaccinated”; “truth about what’s in vaccines”; “vaccines don’t prevent illness”; “how to detox after a vaccine”; “covid vax injury stories”; “covid vaccine doesn’t work”' “refusing the covid shot 2025”; “my body my choice covid vaccine”; “natural immunity is better than vaccine”; “stand against vaccine mandates”; “vaccine choice is human right”; “forced vaccination is tyranny”; “how to avoid vaccine passport”; “doctors speaking out against vaccine”; “banned vaccine truth videos”; “vaccine data they don’t want you to see”; “vaccine linked to health issues”; “vaxxed vs unvaxxed health comparison”; “covid vaccine long term effects”; “5g and vaccine connection?”; “elite agenda behind vaccination”; “watch this vaxxed person still get covid”; “covid cult logic fail”; “vaccine believers getting triggered”; “libs trust Pfizer more than God”; “satire: vaccine commercial parody”; “vaccine-free and healthy”; “how I stayed unvaxxed and safe”; “home remedies over shots” \\ \midrule

climate change support & “proof climate change is real”; “how we know climate change is manmade”; “climate science explained simply”; “what co2 does to the atmosphere”; “climate change effects around the world”; “why we need climate action now”; “climate policy explained”; “debunking climate change misinformation”; “responding to climate denial TikToks”; “how climate myths spread online”; “scientist reacts to climate skeptic”; “climate denier gets schooled”; “climate change disasters 2024”; “melting glaciers time lapse”; “sea level rise explained visually”; “climate crisis is already here”; “how to reduce your carbon footprint”; “climate hope stories”; “gen z fighting for the planet”; “climate deniers be like…”; “funniest climate change denial TikToks”; “watch this guy say co2 is good lol”; “climate denier vs 6th grader”; “climate change explained for conspiracy bros” \\
climate change oppose & “is the earth really getting warmer?”; “global cooling not warming”; “climate change myth explained”; “fake climate crisis exposed”; “evidence earth isn’t warming”; “climate change is natural not manmade”; “earth temperature always changes”; “co2 is not the cause of warming”; “sun cycles explain climate changes”; “climate changes every 100 years”; “climate change agenda exposed”; “why climate policies don’t work”; “climate lockdowns coming?”; “green energy scam”; “world economic forum climate control”; “real scientists against climate change”; “climate truth community”; “banned climate truth videos”; “climate change scam facts”; “volcanoes cause more emissions than people”; “ice caps aren’t melting”; “sea levels not rising proof”; “climate data manipulation exposed”; “how climate activists lie”; “climate cult videos”; “Greta Thunberg meme TikTok”; “climate doomsday predictions gone wrong”; “climate hoax“; “climate change truth seeker“; “no climate change“; “climate change truth“; “is climate change real?“; “climate skeptic“; “stop green deal“; “green deal truth“; “co2 is not a problem“; “is green deal a scam“; “climate wake up“; “no global warming“; “humans do not cause climate change“; “global cooling“; “climate hoax“; “stop climate fearmongering“; “the world is getting colder”; “climate change was a myth” \\ \midrule

political support & "Donald Trump achievements"; "Why Donald Trump is a good president"; "Donald Trump's positive impact on the economy"; "Donald Trump's successful policies"; "Donald Trump's leadership qualities"; "Donald Trump's support for American workers"; "Donald Trump's foreign policy successes"; "Donald Trump's contributions to national security"; "Donald Trump's support for law enforcement"; "Donald Trump's efforts to secure the border"; "Donald Trump's pro-life stance"; "Donald Trump's support for the Second Amendment"; "Donald Trump's fight against fake news"; "Donald Trump's support for religious freedom"; "Donald Trump's commitment to America First"; "Donald Trump's support for small businesses"; "Donald Trump's tax cuts and their benefits"; "Donald Trump's deregulation efforts"; "Donald Trump's support for veterans"; "Donald Trump's energy independence policies" \\
political oppose & "Donald Trump controversies"; "Why Donald Trump is a bad president"; "Donald Trump's negative impact on the economy"; "Donald Trump's failed policies"; "Donald Trump's leadership flaws"; "Donald Trump's lack of support for American workers"; "Donald Trump's foreign policy failures"; "Donald Trump's threats to national security"; "Donald Trump's criticism of law enforcement"; "Donald Trump's border wall failures"; "Donald Trump's anti-immigration stance"; "Donald Trump's opposition to gun control"; "Donald Trump's spread of misinformation"; "Donald Trump's attacks on religious minorities"; "Donald Trump's disregard for international alliances"; "Donald Trump's impact on small businesses"; "Donald Trump's tax cuts and their drawbacks"; "Donald Trump's deregulation and environmental impact"; "Donald Trump's lack of support for veterans"; "Donald Trump's energy policies and climate change" \\ \midrule

cooking & “cooking recipes“; “easy dinner recipes“  \\

\bottomrule

\end{tabularx}
\end{table*}

\begin{table*}[!tbh]
\centering
\small
\caption{Prompt format for the LLM used in the user interaction predictor. The prompt is dynamically constructed based on the topic and available metadata. The parts included in "\{\}" are replaced by the video information, with the \textbf{bolded} parts replaced by the topic or its description as described in this table.}
\label{tab:prompt-format}
\begin{tabularx}{\textwidth}{@{}lX@{}}
\toprule
\textbf{Topic} & \textbf{Prompt Format Text} \\ \midrule
   & \textit{(system part)} You will be provided with an instruction from user regarding video topic and stance annotation. Provide the topic and stance based on the choices and description provided by the user. \\ 
Base prompt & \textit{(user part)} Your task is to determine the topic and the stance of the video given its available metadata. The only possible answers for the topic is: 1) \{\textit{\textbf{topic}}\}; 2) cooking; 3) other. For stance, the possibilities are: 1) support; 2) oppose. Use only these possibilities when answering. \textbackslash n \{\textit{\textbf{topic description}}\} \textbackslash n cooking should be given to anything that is related to recipes or cooking. In this case, the stance does not matter so always put support.\textbackslash n other should be given anything that is not related to previous topics. The stance does not matter so always put support. \textbackslash n Provide the answer in structured form that looks like this: \textbackslash n\textbackslash n Topic: \{\textbf{topic}\}/cooking/other \textbackslash n Stance: support/oppose \textbackslash n\textbackslash n Here is the video information. \textbackslash n Author: \{\textit{author name}\} \textbackslash n Video description and hashtags: \{\textit{video description}\} \textbackslash n Video transcript: \{\textit{voice transcript}\} \textbackslash n Text stickers in video: \{\textit{video stickers}\}  \\ \midrule \midrule

\textbf{Topic} & \textbf{Topic description} \\ \midrule
flatearth & flatearth should be given to anything related to the flatearth conspiracy, including things like firmament or flatearth experiments. The support stance should be given to anything that claims that earth is flat. The oppose stance should be given to anything that claims that earth is not flat, debunks flatearth conspiracy or makes fun of flatearthers.\\ 

vaccines & vaccines should be given to anything related to the vaccines and the discussion on their potential side effects. The support stance should be given to anything that supports vaccinations, motivates people to get vaccinated, debunks vaccine misinformation, acknowledges that there are potential adverse effects in specific cases, acknowledges that you can get covid even if vaccinated, makes fun of people claiming ridiculous side effects or show why people that think vaccines are dangerous do not really understand them. The oppose stance should be given to anything that has anti-vaccination sentiment, claims that vaccines are dangerous for everyone or that vaccines do not work.\\

climate change & climate change should be given to anything related to climate change, global warming or policies that deal with climate change, such as the green deal or reducing co2. The support stance should be given to anything that debunks climate change deniers, claims that we need to changes, shows how the climate change affects us or supports the policies. The oppose stance should be given to anything that denies climate changes, makes fun of it, claims that the earth was always warming, claims there is cooling or opposes the policies for fighting climate change.\\

political & political should be given to anything that deals with politics, elections and other political issues and events in the USA. The support stance should be given to anything related to Trump, republicans, conservatives or any other right leaning content. The oppose stance should be given to anything on the other side of spectrum, such as Biden, Harris, democrats, liberals or anything left learning.\\

\bottomrule

\end{tabularx}
\end{table*}

When choosing the best LLM for the user interaction predictor, we have compared the following models: GPT-4o mini; GPT-4o; LLaMA-3.1-8B; LLaMA-3.1-70B; Gemma-2-9B; Qwen-2.5-7B; GPT-4.1 nano; GPT-4.1 mini; and GPT-4.1 all instruction-tuned variants. In addition, we compare using different prompt formats, where we varied: 1) the metadata included in the prompt (e.g., author, description, transcript, etc.); 2) description for individual topics and stances. The prompt we use for the user interaction predictor is provided in Table~\ref{tab:prompt-format}.

\section{Technical Details of Audit Implementation}
\label{app:technical_details}

\textbf{Account Creation and Location Fidelity.}
To ensure the authenticity of our sock puppet accounts, we implemented comprehensive location masking through Webshare.io's static residential proxy service (100 proxies, 1000 GB monthly bandwidth). During account creation, we configured the FoxyProxy browser extension on local desktop environments and verified through BrowserLeaks.com that our implementation prevented DNS and WebRTC leaks that could expose the true execution environment origin.

Account were created by means of Zoho email services, with each sock puppet configured to match United States localisation parameters, including language preferences, timezone settings, and geographic metadata aligned with the static residential proxy locations. This approach simulated authentic residential user environments characteristic of U.S.-based TikTok users. Within containerised execution environments, we utilised redsocks for SOCKS5 proxy protocol routing.

Early deployment revealed significant challenges with account bans and IP address validation failures. To mitigate these issues, we minimised login frequency and persisted Chrome authentication sessions through user data directories. Each interaction phase execution began with session validation to confirm account status and prevent execution with banned profiles. In addition, prior to the study, we have varied the browser, proxy, timezone and language settings and checked them against various existing bot detectors to validate the users appear authentic for TikTok. This also included preliminary runs of the agents for few days. The settings used as part of the study led to minimal problems with banning - only a single account was banned after the first day (due to the proxy showing incorrect location, e.g., using the same setting, the user was created in the U.S., but appeared to be outside U.S. during interaction), which was quickly replaced by a new one, while there were no other account bans.

\textbf{Browser Automation Architecture.}
Rather than employing conventional automation frameworks (Cypress, Puppeteer, Selenium, NoDriver), we developed a custom agent framework using Chrome DevTools Protocol (CDP). This design decision provided several advantages: reduced automation footprint detection, faster adaptation to platform changes, direct access to media events (video playback state, buffering), and real-time network traffic monitoring. The implementation avoided common automation signatures such as missing audio services or simulated display contexts that trigger platform detection mechanisms.

\textbf{Video Identification and Content Tracking.}
Video identification required sophisticated network traffic analysis of multiple TikTok API endpoints (recommendation, search, feed cache) to construct a comprehensive view of available content. Since the video presentation order did not correspond to the network response sequence, we implemented a matching algorithm correlating displayed author names and descriptions with network payload metadata to identify the currently playing video. This approach enabled accurate retrieval of video IDs for subsequent interaction prediction requests.

\textbf{Interaction Prediction System.}
The interaction prediction service operates as a FastAPI application with 2 Uvicorn workers per node, supporting 2 concurrent agents (1:1 agent-to-predictor ratio). For each displayed video, the service received user profile attributes (age, gender, topic preference, stance), video metadata (author, description), and audio transcription generated via the Whisper-3-large-turbo model. The service returned interaction decisions: bookmark, like, watch, or skip.

To optimise resource utilisation, we implemented Redis-based caching for both transcriptions and user reactions, substantially reducing redundant processing during the seed phase where users encountered overlapping content. All predictor deployments shared a single Redis instance, enabling cross-agent cache utilisation. At peak operation, we maintained 8 predictor instances serving 16 concurrent agents.

\textbf{Infrastructure and Orchestration.}
The experimental infrastructure deployed on AWS EKS with heterogeneous node types: t3a.xlarge and c6a.large for general computation, r6a.large for memory-intensive operations, and g4dn.xlarge GPU instances for interaction predictors. Apache Airflow orchestrated agent scheduling and execution coordination, while Prometheus and Grafana provided metrics collection and operational visibility.

Each agent session generated comprehensive interaction logs stored as Parquet files containing timestamped user actions and video metadata. Kubernetes Persistent Storage isolated browser sessions and collected data per agent, with final artefacts transferred to S3 for archival storage.

\textbf{Challenges and Mitigation Strategies.}
Despite the 10-day study duration, maintaining operational stability presented challenges. Account bans occurred sporadically, requiring ongoing monitoring and intervention. TikTok's evolving web application structure necessitated selector updates as HTML markup changed. Novel pop-up dialogues and CAPTCHA challenges interrupted execution flows in a subset of runs.

Initial login procedures proved most problematic, often requiring multiple attempts or resulting in account bans. We deployed SadCaptcha for automated CAPTCHA solving and maintained VNC remote desktop access for manual intervention during critical authentication failures. Proxy rotation by the provider occasionally resulted in country-of-origin mismatches, triggering platform detection and subsequent bans. We addressed this through pre-execution proxy validation, verifying location consistency before each interaction phase to prevent mid-session origin changes.

Once agents successfully entered the interaction phase, execution stability improved substantially, with issues primarily limited to proxy infrastructure changes rather than platform interaction failures. This operational profile informed our retry logic and checkpoint strategies, concentrating resilience mechanisms around the authentication boundary.

\section{Supplementary Results Details}
\label{app:supplementary-results}

For the Flatearth topic, we observe the following distribution of recommended videos over the whole audit: $6822$ unrelated, $472$ topic relevant, and $2675$ cooking.

For the Vaccination topic, we observe the following distribution of recommended videos over the whole audit: $9372$ unrelated, $121$ topic relevant, and $510$ cooking.

For the Climate Change topic, we observe the following distribution of recommended videos over the whole audit: $3894$ unrelated, $3$ topic relevant, and $1202$ cooking.

For the US politics topic, we observe the following distribution of recommended videos over the whole audit: $3953$ unrelated, $3010$ topic relevant, and $714$ cooking.

\section{Topic-Related Hashtag Popularity Analysis}

To determine whether the observed behaviour can be explained by the popularity of different topics, we run further analysis. For each topic, we extract the hashtags from videos that are watched during the seed phase that are related to the topic (e.g., we select all the videos that are related to the flatearth topic that were watched during the seed phase and extract hashtags from them). We count the occurrences of these hashtags across videos and select the 10 most common ones, while removing the generic ones not related to the topic (such as "fyp"). We determine the popularity for each of these hashtag based on the overall number of videos that belong to it (based on the TikTok hashtag search\footnote{Can also be determined through this page \url{https://ads.tiktok.com/business/creativecenter/inspiration/popular/hashtag/pc/en}}) and sum it across all such hashtags. The final topic popularity results are: 177600000 views for cooking, 22330600 for vaccines, 31139700 for US politics, 1197900 for climate change and 593700 for flatearth.

Based on this, we can observe that the cooking and US politics topics have similar popularity, followed by flatearth and vaccines (again with similar popularity) and finally the climate change topic. However, we observe significantly different behaviour regarding the individual topics that share similar popularity. As such, we can determine that the popularity is not the sole reason for the different behaviour.

\end{document}